\documentclass{article}

\usepackage[english]{babel}

\begin{document}

\title{Child Universes in the Laboratory}

\date{{\footnotesize{}November 5th, 2006}}

\author{Stefano Ansoldi\footnotemark[1]\\
{\small{}International Center for Relativistic Astrophysics (ICRA)}\\
{\small{}and Universit\`{a} degli Studi di Udine, Udine, Italy}\\
{\small{}\textrm{email:} \texttt{ansoldi@trieste.infn.it}}\\[1mm]
Eduardo I. Guendelman\\
{\small{}Ben Gurion Univeristy, Beer Sheva, Israel}\\
{\small{}\textrm{email:} \texttt{guendel@bgu.ac.il}}
\footnotetext[1]{Mailing address: Dipartimento di Matematica e Informatica,
Universit\`{a} degli Studi di Udine, via delle Scienze 206,
I-33100 Udine (UD), Italy}}

\maketitle

\begin{abstract}
Although cosmology is usually considered an
observational science, where there is little
or no space for experimentation, other approaches
can (and have been) also considered. In particular,
we can change rather drastically the above, more
passive, observational perspective and
ask the following question: could it be possible,
and how, to create a universe in a laboratory?
As a matter of fact, this seems to be possible,
according to at least two different paradigms;
both of them help to evade the consequences
of singularity theorems.
In this contribution we will review some of these models
and we will also discuss possible extensions
and generalizations, by
paying a critical attention to the still open issues as,
for instance, the detectability of child universes and the
properties of quantum tunnelling processes.

\end{abstract}

\section{The studies so far \dots{}}

The world Cosmology stems from the Greek word \textit{cosmos},
which meant \textit{beauty}, \textit{harmony}, and is the name
of that branch of science which studies the origin and evolution
of the universe. Thus, considering its name and the object of its study,
it is, perhaps, natural to take a ``passive'' point of
view when dealing with cosmological problems, where we use
the word \emph{passive} to emphasize that our experience of cosmology
is mainly observational in nature. This may undoubtedly
be a condition that seems hard to change \emph{in practice}: after
all we are dealing with problems, as the birth
of our universe and its evolution in the present
state, which do not appear suitable for a direct experimental approach.
On the other hand, we do not see any reason why this should prevent
us from changing our \emph{attitude} toward the problem, switching
from a contemplative to a more active one.
In our opinion, a stimulation in this sense is coming already
from the theory which first gave
us the opportunity to address cosmological
problems quantitatively, i.e.
General Relativity. General Relativity
raises for the first time the concept of
causality as a central one in physics.
This means that, taking a very pragmatic
point of view, we have to
admit that only a subset of what exists in
our universe can be experienced/observed
by us. This is not because of our limited
capabilities as humans, but, more
fundamentally, because of the restrictions
imposed by the spacetime structure
on the causal relations among objects.
At the same time causality also brings a challenge
to cosmologists in connection with the \emph{large scale} spacetime structure;
this is because the simplest models
of the universe which are built according to General Relativity
and whose late time predictions have a reasonable degree of consistency with what we observe,
seem doomed to have an initial singularity in their past so that field equations
break down exactly where we would like to set up the initial conditions.
This undesirable situation looks even more disappointing after the observation
that many parameters describing the state of the early universe
are quite far from the
domain of ``very large scales'' which characterizes
the present observable universe. Let us, for instance,
consider
a Grand Unified Theory scale of
$10 ^{14}$ Gev: the universe could then emerge
from a classical bubble which starts
from a very small size and has a mass of
of the order of about $10$ Kg (by using quantum tunnelling
the mass of the bubble could be
arbitrarily small, but the probability of
production of a new universe out
of it would be reduced).
The density of the universe would, admittedly,
have been quite higher than
what we could realize with present technologies,
but the orders of magnitude of the
other parameters make not unreasonable to
ask the question: might we have the possibility
of building a child universe in the laboratory?

As a matter of fact,
a positive answer to this question was already
envisaged some years ago (for a popular level
discussion see \cite{bib:NewSc20062559....32M}).
In particular Farhi \emph{et al.} suggested an interesting model able
to describe universe creation starting from a
non-singular configuration and involving
semiclassical effects. This proposal, actually,
leaves some open issues, for instance about
the semiclassical part of the process and the
global (Euclidean) structure of the solution.
Although since then, a few more proposal
have appeared, addressing in more detail
qualitative issues, it is interesting to observe
that most of the problems which emerged
in the earliest formulation are,
somehow, still open. It is our hope that
the present review of the different approaches
which have been developed along this interesting
research line, will stimulate to study
in more detail and with systematic
rigor these problems as well as other
realistic answers to the above question.
In our opinion, this question is not a purely
academic one, and might help not only to change
our perspective (passing from an observational
to an experimental one) in addressing cosmological
problems, but also to shed some light on the importance
of the interplay of gravitational and quantum phenomena.

We would also like to remark how, this complementary
perspective can be considered much more promising nowadays
than some years ago, thanks to the results of recent observations.
These observations are helping us in focusing our field of view
back in time, closer and closer to the earliest stages of life
of our universe and are providing us with a large amount of data
and information that will, hopefully, help us in sharpen our
theoretical models. This has already
allowed tighter constrains on the parameters of
models of the early universe, giving us the chance
for a more decisive attack of most of the still open
problems. This will be a great help also for a
``child universe formation in the laboratory''
program; it can make easier to identify the
fundamental elements (building blocks)
required to model the \emph{creation} of a
universe that will evolve in something similar to
the present one. At the same time, it will help
us to narrow our selection of the fundamental
principles that forged the earliest evolution of
the universe, and, as we said already before,
strengthen our hope to enlighten
a crucial one, which is the interplay between
General Relativity and Quantum Theory.

This said, in the rest of this section, keeping in mind the
above preliminary discussion, we are going
to give a concise review of the state of the
art in the field and to to make a closer contact
with some of the models for child universe formation;
in particular we are going to review some of the existing
works on the dynamics of vacuum bubbles and on
topological inflation
(both also considered
in a semiclassical framework).

Callan and Coleman initiated the study of vacuum decay more than
30 years ago \cite{bib:PhReD1977..15..2929C,bib:PhReD1977..16..1762C};
after their seminal papers the interest in the subject rapidly increased.
The possible interplay of true vacuum bubbles with gravitation was then
considered \cite{bib:PhReD1980..21..3305L,bib:PhReD1987..36..1088W}.
More or less at the same time and as opposed with the true vacuum
bubbles of Coleman \textit{et al.}, false vacuum bubbles were also considered.
The classical behavior of
regions of false vacuum coupled to gravity
was studied by Sato et \textit{al.}
\cite{bib:PrThP1981..65..1443M,%
bib:PrThP1981..66..2052M,bib:PrThP1981..66..2287S,%
bib:PhLeB1982.108....98K,bib:PhLeB1982.108...103M,%
bib:PrThP1982..68..1979S} and followed by the works
of Blau \textit{et al.} \cite{bib:PhReD1987..35..1747G}
and Berezin \textit{et al.} \cite{bib:PhReD1987..36..2919T,%
bib:PhReD1991..43.R3112T}. The analysis in \cite{bib:PhReD1987..35..1747G}
clarified some aspects in the study of false vacuum
dynamics coupled to gravity; in it, for the first time, the problem
was formulated using geodesically complete coordinate systems:
this made more clear the issue of wormhole formation,
with all its rich sequel of stimulating properties and consequences.

The presence of wormholes makes possible a feature of false vacuum
bubbles that is otherwise counterintuitive, which is
that these objects can undergo an exponential inflation
without displacing the space outside of the bubble itself;
this could seem strange at a first look and is due to the fact
that they have an energy density which is higher
than that of the surrounding spacetime and which is
responsible for keeping the required pressure difference.
Because of this, \emph{child universe} solutions appear
as expanding bubbles of false vacuum
which disconnect from the exterior region.
Apart from the already mentioned wormhole, they are also
characterized by the presence
of a \emph{white-hole like} initial singularity;
the simplest example can be obtained by modelling
the region inside the bubble with a
domain of de Sitter spacetime and the
region outside the bubble with a domain of Schwarzschild spacetime.
These two regions are then joined across the bubble
surface, using the well known Israel junction conditions
\cite{bib:NuCim1966.B44.....1I,%
bib:PhReD1991..43..1129I}; Einstein equations, which hold
independently in the two domains separated by the bubble, are
also satisfied on the bubble surface if interpreted in a distributional
sense; they determine the motion (embedding)
of the bubble in the two domains of spacetime.
Although there are various simple configurations of this system
(as well as more elaborate generalizations)
that are appropriate to describe the evolution
of a newly formed universe (i.e. they are such that
the expanding bubble can become very large),
these \emph{classical} models present also some
undesirable features.
    In particular it turns out that
    only bubbles with masses above
    some critical value can expand from very small
    size to infinity.
    But then these solutions necessarily have a (white-hole) singularity
     in their past; in fact, for all of them the
    hypotheses of singularity theorems are satisfied.

In connection with the restriction on the values of the total mass,
the situation could be improved in theories containing an appropriate multiplet
of additional scalars\cite{bib:PhReD1991..44..3152R}\footnote{{The
subject of inflation assisted by topological
defects was also studied later in
\cite{bib:PhReL1994..72..3137V} and
\cite{bib:PhLeB1994..327...208L}.}}:
then all bubbles that start evolving from zero radius can inflate
to infinity if the scalars are in
a ``hedgehog'' configuration, or global monopole
of big enough strength. This effect also
holds in the gauged case for magnetic monopoles
with large enough magnetic charge: in this way the
mass requirement is traded for requirements about the properties
of magnetic monopoles.

A possible connection of this approach with the problem of the
initial singularity appears, then, from the work
of Borde \textit{et al.}
\cite{bib:PhReD1999..59043513V}: they
proposed a mechanism which,
by means of the coalescence of two regular magnetic
monopoles (with \textit{below critical}
magnetic charge), is able to produce
a \textit{supercritical} one, which then
inflates giving rise to a child universe.
This idea might help addressing the
singularity problem and in this context it is very
interesting the work of Sakai \textit{et al.}
\cite{bib:gr-qc2006..02...084K}: in it
the interaction of a magnetic monopole
with a collapsing surrounding membrane
is considered; also in this case a
new universe can be created and the presence
of an initial singularity \emph{in the causal past
of the newly formed universe} can be avoided.

To solve the problem of initial singularity, there
are also other approaches which make a good use of
quantum effects. Needless to say, these ideas are
very suggestive because they require a proper interplay
of quantum and gravitational physics, for which a consistent
general framework is still missing. This is the main reason
why most of these investigations try to obtain a simplified
description of the system by requiring a high degree of
symmetry from the very beginning. In particular, if
we describe the bubble separating the inflating
spacetime domain from the surrounding spacetime
in terms of Israel junction conditions
\cite{bib:NuCim1966.B44.....1I,%
bib:PhReD1991..43..1129I}, under the additional assumption
of spherical symmetry, the dynamics of the system is
determined by the dynamics of an effective system with only one degree
of freedom: this is called the \textit{minisuperspace
approximation}; in this framework the problem
of the semiclassical quantization of the system, even
in the absence of an underlying
quantum gravity theory, can be undertaken with less (but
still formidable) technical problems using as a direct guideline
the semiclassical procedure with which we are familiar in ordinary
Quantum Mechanics. This has been the seminal idea of
Farhi \textit{et al.} \cite{bib:NuPhy1990B339...417G} and of Fishler \textit{et al.}
\cite{bib:PhReD1990..41..2638P,bib:PhReD1990..42..4042P}.
One additional difficulty in these approaches was in connection with the
stability of the classical initial state. Interestingly enough,
this could be solved by the introduction of massless scalars or gauge fields
that live on the shell and produce a classical stabilization effect
of false vacuum bubbles.
By quantum tunnelling, these bubbles can then become child universes
\cite{bib:ClQuG1999..16..3315P} and, at least in a $2+1$-dimensional example
\cite{bib:MPhLA2001..16..1079P}, it has been shown that
the tunnelling can be arbitrarily small.

\section{\dots{} and their future perspectives}

From the above discussion, we think it is already clear that there
are many interesting aspects in the study of models for child universe
creation in the laboratory. We would also like to remember how
most of these models are based on a very well-known
and studied classical system,
usually known as a \textit{general relativistic shell}
\cite{bib:NuCim1966.B44.....1I,%
bib:PhReD1991..43..1129I}. The classical dynamics
of this system is thus ``under control'',
many analytical results can be found in the literature
and numerical methods
have also been employed (see the introduction of
\cite{bib:ClQuG2002..19..6321A} for
additional references). On the other hand there has been
little progress in the development of the quantized
theory, which still remains a
non-systematized research field. We stress how a progress in
this direction would be decisive
for a more detailed analysis of the
semiclassical process of universe creation.

Before coming back to the quantum side of the problem,
let us first consider what could be done on
the classical one. We will concentrate mainly
on the works
of Borde \textit{et al.} \cite{bib:PhReD1999..59043513V}
and of Sakai \textit{et al.} \cite{bib:gr-qc2006..02...084K},
which suggest many interesting ideas for further
developments.
    For instance, it is certainly important to extend
    the analysis in \cite{bib:PhReD1999..59043513V}, which is
    mainly qualitative in nature, to take fully into account
    the highly non-linear details of the collision process
    by means of which a supercritical monopole is created
    (this is certainly instrumental for a quantitatively
    meaningful use of the idea of topological inflation).
    Also the study performed in \cite{bib:gr-qc2006..02...084K}
    should be extended; to obtain some definitive conclusion about
    the stability of the initial configuration, it is, in fact,
    necessary to study the spacetime structure of the model for
    all possible values of the parameters; it could then be possible
    to determine if stability is a general feature of
    monopole models or an \emph{accident} of some particular
    configurations.
From the classical point of view, in both the above models another
central point is the study of their causal structure; it can be
obtained by well-known techniques, but, again, a full classification
of all the possibilities that can arise
is certainly required to gain support for the proposed mechanisms.
Known subtleties which require closer scrutiny
(as for example, the presence of singularities
in the causal past of the created universe
\emph{but} not in the past of the experimenter
creating the universe in the laboratory or, sometimes,
the presence of timelike \emph{naked} singularities)
make a discussion of the
problem of initial conditions not only interesting
but necessary, especially in this context\footnote{The proper
analysis of the Cauchy problem will, in fact,
involve resolution or proper handling of these
singularities.}.

A suggestive complement to the classical aspects discussed
above, is represented, of course, by the quantum
(more precisely semiclassical) ones, where
quantum effects are advocated to realize
the tunnelling between classical solutions. If
(i) the classical solution used to describe the initial state can
be formed without an initial
singularity and is stable, (ii) the classical solution which represents the final state
can describe an inflating universe and (iii) we can master
properly the tunnelling process, then we could
use the quantum creation of an inflating
universe \emph{via} quantum tunnelling
to evade the consequences of singularity
theorems. The construction of proper initial and final states
has already been successfully accomplished. The
stability of the initial classical configuration has been,
instead, only partly analyzed \cite{bib:ClQuG1999..16..3315P}
and it would be certainly interesting to
consider the tunnelling process in more
general situations, where, for example,
the stabilization can be still classical
in origin. Although there is some evidence
\cite{bib:gr-qc2006..02...084K}
of a general way to solve this issue
in the context of monopole
configurations, as we mentioned
above, the analysis should be extended
to the whole of the parameter space.
At the same time a complementary possibility is
that \emph{semiclassical}
effects might stabilize the initial
configuration. In particular,
closely related to the problem of
instabilities present in many models,
is the fact that the spacetime surrounding the vacuum
bubble has itself an instability due to presence of a
white hole region (see, for instance,
\cite{bib:PhReD1977..16..3359T}). Also in this
context quantum effects might stabilize the system
and help solving the issue. This approach could
require the determination of the
stationary states of the system in the
WKB approximation, a problem for which
a generalization of the procedure presented in
\cite{bib:ClQuG2002..19..6321A} (where
this analysis was performed for the first
time in a simplified model) could be useful.

Another equally (if not more) important point for future
investigations is certainly related with the still
open issues in the semiclassical tunnelling procedure.
We will shortly discuss this
by following, for definiteness,
the clear, but non-conclusive,
analysis developed by Farhi
\textit{et al.} \cite{bib:NuPhy1990B339...417G}:
it is shown in their paper that, when considering
the tunnelling process, it is not possible
to devise a clear procedure to build the manifold
interpolating between the initial and final
classical configurations; this
manifold would describe the instanton that is
assumed to mediate the process. According to
the discussion of Farhi \textit{et al.}
it seems possible to build
only what they call a \emph{pseudo-manifold},
i.e. a manifold in which various points
have multiple covering. To make sense of
this, they are forced to introduce
a `covering space' different from the
standard spacetime manifold, in which
they allow for a change of sign of the
volume of integration required for
the calculation of the tunnelling action
and thus of tunnelling probabilities.
It would be important to put on a more solid
basis this interesting proposal, comparing it
with other approaches
which might help to give a more precise
definition of this
\emph{pseudo-manifold}. In particular
we would like to mention two possibilities.
A first one uses the \textit{two measures
    theory} \cite{bib:PhReD1999..60065004K}; considering
    four scalar fields it is possible to define
    an integration measure in the action
    from the determinant of the mapping between
    these scalar fields and the four
    spacetime coordinates; there can, of course, be
    configurations where this mapping is not of
    maximal rank and if we then interpret the scalar
    fields as coordinates
    in the \emph{pseudo-manifold}
    of \cite{bib:NuPhy1990B339...417G}, then the
    non-Riemannian volume
    element of the two measures theory would be
    related to the non-Riemannian
    structure that could be associated to the
    \emph{pseudo-manifold}. In this perspective,
    non-Riemannian
    volume elements could be essential to make
    sense of the quantum creation of a universe in
    the laboratory and it could be important to develop
    the theory of shell dynamics in the framework described
    by the two measures theory.

A second one, likely complementary,
    can come from a closer study of the Hamiltonian
    dynamics of the system.
    Let us preliminarily remember
    that the Hamiltonian for a general relativistic
    shell, which we are using as a model for the
    universe creation process, is a non quadratic
    function of the momentum (this comes from the
    non-linearities intrinsic to General Relativity);
    this makes the quantization procedure non-standard
    and quite subtle too. Moreover, although it is possible
    to determine an expression for the Euclidean
    momentum and use it to reproduce \cite{bib:ClQuG1997..14..2727A}
    standard results for the decay of vacuum bubbles (as for instance
    the results of Coleman \emph{et al.} \cite{bib:PhReD1980..21..3305L})
    this momentum can have unusual properties
    along the tunnelling trajectory; some of these inconsistencies
    disappear if we consider the momentum as
    a function valued on the circle instead than on
    the real line \cite{bib:gr-qc2006..xx...zzzS} but
    further investigations in this direction
    are required; they will likely help us to obtain
    a better understanding of the semiclassical tunnelling creation
    of this general relativistic system and, perhaps, show us
    some interesting properties of the interplay between the quantum
    and the gravitational realms. In this context, it should be also explored how the
    Euclidean baby universes \cite{bib:Proc.1989Pakistan...N} could be matched
    continuously to the real time universes and in this way provide new
    ways to achieve spontaneous creation of real time baby universes

To complement the above discussion, we would now like to provide some
additional contact points between theoretical
ideas and experimental evidence. We start
considering if all \emph{creation efforts} might end
in a child universe totally disconnected from its
\emph{creator} or not. Of course, there is not a definitive
answer also to this problem yet, since this is tightly bound
to the child universe creation model. Nevertheless, it
it is certainly stimulating to address
the question if, in some way, the new
universe might be detectable. There is an
indication in this direction from the
analysis performed in
\cite{bib:PhReD1991..44...333M}: here a
junction with a Vaidya radiating metric is employed,
so that the child universe could be detectable because
of modifications to the Hawking radiation.
Generalizations that apply
to solitonic inspired universe creation%
\footnote{It is, for instance, certainly possible to extend
the metric describing
the monopole, i.e.
the Rei\ss{}ner-Nordstr\"{o}m spacetime, to the
Rei\ss{}ner-Nordstr\"{o}m-Vaidya case.}
can be important, especially from the point of view
of a quantum-gravitational scenario in which the
exact and definite character of classical
causal relations might be \emph{waved}
by quantum effects.

Other issues that could be tackled after having
a more detailed model of child universe
creation, are certainly phenomenological ones.
They would also help to better understand the
differences between purely classical and partly quantum
processes, which is also a motivation to consider
them explicitly and separately. Also
the physical consequences of different values of
the initial parameters characterizing the child
universe formation process (initial conditions)
should be analyzed\footnote{In particular different
ways of creating a universe
in the laboratory could lead to different
coupling constants, gauge groups, etc..}
and in this context we would also like to recall the
idea of Zee \textit{et. al}
\cite{bib:phys.2005..10...102Z}, i.e. that a creator
of a universe could pass a message to the future
inhabitants of the created universe. From our
point of view this is can be a suggestive
way to represent the problem
of both initial conditions and
causal structure; this could be of relevance
also for the problem of defining probabilities
in the context of the multiverse theory and
of eternal inflation.

A final point of phenomenological relevance
would be in connection with observations that suggest
the universe as super-accelerating. This seems to
support the idea that some very unusual
physics could be governing the universe,
in the sense that standard energy conditions
might not be satisfied. In the context of
child universes creation in the laboratory
in the absence of an initial singularity,
it might very well be
that a generalized behavior of the universe to
try to raise its vacuum energy would manifest
itself locally with the
creation of bubbles of false vacuum (as seen
by the
surrounding spacetime), which would then led
to child universes. In \cite{bib:gr-qc2006..07...111G}
a proposal, based on the two
measures theory, to avoid initial
singularities in a homogeneous cosmology has already
been put forward. It would then be
desirable to apply it to the
non-singular child universe creation also.

To conclude we cannot miss to point out how
all the above discussion about the possibility
of producing child universes in the laboratory
could take a completely new and concrete perspective
in connection with the possible existence of new physics
at the TeV scale in theories with large
compact extra-dimensions, physics that might become
available to our experimental testing at the
colliders which will shortly start to operate.

\section*{Acknowledgements}

We would like to thank H. Ishihara and
J. Portnoy for conversations.

\end{document}